\magnification=\magstep1 
\font\bigbfont=cmbx10 scaled\magstep1
\font\bigifont=cmti10 scaled\magstep1
\font\bigrfont=cmr10 scaled\magstep1
\vsize = 23.5 truecm
\hsize = 15.5 truecm
\hoffset = .2truein
\baselineskip = 14 truept
\overfullrule = 0pt
\parskip = 3 truept
\def\frac#1#2{{#1\over#2}}

\nopagenumbers
%
\topinsert
\vskip 3.2 truecm
\endinsert
\centerline{\bigbfont RELATIVISTIC THEORY OF SUPERCONDUCTIVITY}
\vskip 20 truept
\centerline{\bigifont K. Capelle}
\vskip 8 truept
\centerline{\bigrfont Departamento de Qu\'{\i}mica e F\'{\i}sica Molecular}
\vskip 2 truept
\centerline{\bigrfont Instituto de Qu\'{\i}mica de S\~ao Carlos, USP}
\vskip 2 truept
\centerline{\bigrfont 13560-970 S\~ao Carlos, SP, Brazil} 
\vskip 14 truept
\centerline{\bigifont M.A.L. Marques}
\vskip 8 truept
\centerline{\bigrfont Departamento de F\'{\i}sica Te\'{o}rica}
\vskip 2 truept
\centerline{\bigrfont Faculdad de Ciencias, 47011 Valladolid, Spain}
\vskip 14 truept
\centerline{\bigifont E.K.U. Gross}
\vskip 8 truept
\centerline{\bigrfont Institut f\"ur Theoretische Physik}
\vskip 2 truept
\centerline{\bigrfont Universit\"at W\"urzburg, D-97074 W\"urzburg, Germany}
\vskip 1.8 truecm

\centerline{\bf 1.  RELATIVISTIC ORDER PARAMETERS}
\vskip 12 truept
Relativistic effects in materials containing heavy atoms pervade many areas of 
condensed-matter physics [1-4]. 
The present chapter is devoted to the theory of relativistic effects in a
particularly interesting class of materials, namely
superconductors and other superfluids [5-9].
At its present stage that theory is limited to single-particle effects
(on the level of the Dirac equation); a relativistic description of the
particle-particle interaction is not attempted. However, already on
the single-particle level several unexpected relativistic phenomena in
superconductors emerge, among them a novel type of spin-orbit
coupling, present in superconductors only. The origin and the physics of
these phenomena are discussed in this contribution.

The fundamental ingredient in any description of superconductivity
is the order parameter (OP). The 
BCS theory of superconductivity [10], in its original formulation, 
is not easily generalized to the relativistic domain.
Effects of spatial inhomogeneity of the lattice and the OP,
of different OP symmetries, and of magnetic fields are, however,
easily incorporated within the framework of the Bogolubov-de~Gennes (BdG) 
reformulation of the BCS theory [11], which also provides a suitable
basis for the relativistic generalization. In the conventional BdG theory
the OP describing the Cooper pairs can be written in terms of 
products of field operators,
$\hat{\psi}_\uparrow(r)\hat{\psi}_\downarrow(r')$, 
$\hat{\psi}_\downarrow(r)\hat{\psi}_\uparrow(r')$, 
$\hat{\psi}_\uparrow(r)\hat{\psi}_\uparrow(r')$, and
$\hat{\psi}_\downarrow(r)\hat{\psi}_\downarrow(r')$,
where the $\hat{\psi}_\sigma$ are components of the two-component Pauli
spinor $\hat{\Psi}$. (All spatial arguments $r$ are actually vectors $\vec{r}$,
but for ease of notation the arrows are suppressed here and below.)
Linear combinations of these products can be formed to represent the usual 
singlet and triplet 
states, which are even or odd under interchange of $r$ and $r'$,
respectively [6,11-14]. The singlet pair is described, for example, by
$\hat{\Psi}^T(r)i\hat{\sigma}_y\hat{\Psi}(r') = 
\hat{\psi}_\uparrow(r)\hat{\psi}_\downarrow(r') -
\hat{\psi}_\downarrow(r)\hat{\psi}_\uparrow(r')$,
which is manifestly even under interchange of $r$ and $r'$, and odd under
that of spin up and spin down, as expected for a singlet state.
Since the Pauli matrix $i\hat{\sigma}_y$ is the nonrelativistic 
time-reversal matrix, this way of writing the pair additionally brings out 
clearly that the BCS Cooper pair consists of time-conjugate single-particle
states [11,13]. 

A convenient way of describing 
arbitrary pair states is by expanding the general $2\times2$ pair potential 
matrix $\hat{\Delta}(r,r')$ in the set of matrices $i\hat{\sigma}_y$ and 
$i\vec{\sigma}\hat{\sigma}_y$, where $\vec{\sigma}$ is the vector of Pauli 
matrices, and the three
components of $i\vec{\sigma}\hat{\sigma}_y$ are the matrices appearing in
the Balian-Werthamer parametrization of triplet OP [15], 
$$
\hat{\Psi}^T(r)\hat{\Delta}(r,r')\hat{\Psi}(r') =
\Delta_0(r,r')\hat{\Psi}^T(r)i\hat{\sigma}_y\hat{\Psi}(r')
+\vec{\Delta}(r,r')\hat{\Psi}^T(r)i\vec{\sigma}\hat{\sigma}_y\hat{\Psi}(r').
\eqno(1)
$$
Ordinarily one would use the unit matrix $\hat{I}$ and the vector of Pauli 
matrices $\vec{\sigma}$ as basis for this type of expansion, since by using 
this symmetry adapted set of matrices one achieves a separation of scalar
quantities ($\hat{\Psi}^\dagger \hat{I}\hat{\Psi}$) from quantities 
transforming as 
a vector ($\hat{\Psi}^\dagger \vec{\sigma}\hat{\Psi}$), i.e, a classification
with respect to irreducible representations (IR) of the rotation group. 
However, Cooper pairs are formed with two creation (or two annihilation) 
operators, and not with one creation and one annihilation operator, and a 
scalar-vector separation is only achieved in terms of the Balian-Werthamer 
matrices [15]. ($\hat{\Psi}^T \vec{\sigma}\hat{\Psi}$, for example, does not 
transform as a vector under rotations, while 
$\hat{\Psi}^Ti\vec{\sigma}\hat{\sigma}_y\hat{\Psi}$ does.)

In a relativistic formulation the spin index $\sigma$ is replaced by the label 
for the components of four-component Dirac spinors. Out of these four 
components one can form sixteen pairs, which can again be represented
in terms of a complete set of matrices. One might be tempted to employ the 
sixteen $4\times4$ Dirac $\hat{\gamma}$ matrices for this purpose, since these
usually lead to a separation into so called `bilinear covariants', i.e.,
objects transforming under some IR of the Lorentz group
${\cal L}$ [16]. However, just as in the nonrelativistic case, the usual set
of matrices does not achieve this if the object to be expanded is formed
with two field operators of the same kind. A set of matrices in terms of
which such a separation into IR of ${\cal L}$ is achieved, was derived in
Ref.~[6], where it was shown that the usual classification into five bilinear
covariants, a Lorentz scalar (one component), a four vector (four 
components), a pseudo scalar (one component), an axial four vector (four
components), and an antisymmetric tensor of rank two (six independent
components), carries over to the superconducting OP if the Dirac $\hat{\gamma}$ 
matrices are replaced by a new set of matrices, denoted $\hat{\eta}$.
Explicit expressions for these $\hat{\eta}$ matrices are given in 
Ref.~[6]. These five OP, with a total of sixteen components, exhaust the 
possible pairings which can be formed from two Dirac spinors.
As an example, consider the most important OP of all, the BCS singlet OP,
which can be written in terms of Pauli spinors as 
$\hat{\chi}^{\rm BCS,non-rel}=\hat{\Psi}^Ti\hat{\sigma}_y \hat{\Psi}$. 
Its relativistic generalization reads
$\hat{\chi}^{\rm BCS,rel} = \hat{\Psi}^T\hat{\eta}_0\hat{\Psi}$, where 
$\hat{\eta}_0=\hat{\gamma}^1\hat{\gamma}^3$ is one of the sixteen $\hat{\eta}$
matrices (the one yielding a Lorentz scalar), the $\hat{\Psi}$ are now 
four-component Dirac operators, and $\hat{\gamma}^1$ and $\hat{\gamma}^3$ are 
two of the standard Dirac matrices.

Initially [5], the relativistic generalization of the BCS singlet OP
was constructed by replacing the nonrelativistic time-reversal matrix 
$\hat{t}=i\hat{\sigma}_y$ in $\hat{\chi}^{\rm BCS,non-rel}$ by the 
relativistic one, $\hat{T}=\hat{\gamma}^1\hat{\gamma}^3$, when constructing 
$\hat{\chi}^{\rm BCS,rel}$, 
i.e., it was postulated that relativistic Cooper pairs still consist of
time-conjugate states. In later work [6], this turned out to be unnecessary,
since the matrix $\hat{\eta}_0\equiv \hat{T}$ automatically emerges as that of 
the sixteen $\hat{\eta}$ matrices leading to covariant pairs, which
reduces to the BCS singlet OP in the nonrelativistic limit. It is important
to point out that the relativistic generalization of the BCS OP is uniquely
determined by both prescriptions, and that both indeed lead to the same
result, $\hat{\chi}^{\rm BCS,rel} = \hat{\Psi}^T\hat{\eta}_0\hat{\Psi}$.

\vskip 28 truept
\centerline{\bf 2.  DIRAC EQUATION FOR SUPERCONDUCTORS}
\vskip 12 truept

In terms of the relativistic OP one can now generalize the entire theory of
superconductivity to the relativistic domain. 
At the heart of this generalization are the Dirac-Bogolubov-de~Gennes
equations (DBdGE), which for a BCS-like spin-singlet superconductor read [5,6]
$$
\hat{\gamma}^0
[c \vec{\gamma} \cdot \vec{p} + mc^2(1-\hat{\gamma}^0)+q\gamma^\mu A_\mu]
u_{n}(r)  
+ \int d^3r'\Delta(r,r')\hat{\eta}_0 v_{n}(r')
= E_{n}  u_{n}(r)
\eqno(2)
$$ 
$$
-\hat{\gamma}^0
[c \vec{\gamma} \cdot \vec{p} + mc^2(1-\hat{\gamma}^0)+q\gamma^\mu A_\mu]^*
v_{n}(r) 
-\int d^3r' \Delta^*(r,r')\hat{\eta}_0 u_{n}(r')
= E_{n} v_{n}(r),
\eqno(3)
$$
where both $u_{n}$ and $v_{n}$ are four-component (Dirac) spinors, representing
particle and hole amplitudes, while $\gamma^\mu$ is the four vector of 
$\hat{\gamma}$ matrices (in standard notation, and with a summation over 
repeated greek indices implied), and $\vec{\gamma}$ the corresponding three 
vector, containing $\hat{\gamma}^1$,$\hat{\gamma}^2$, and $\hat{\gamma}^3$.
$A_\mu$ is the four potential, and $\Delta$ the pair potential.
Equations (2) and (3) generalize the conventional Bogolubov-de~Gennes equations
(BdGE) for spin-singlet superconductors [11] to the relativistic domain [5,6] 
in the same way in which the conventional Dirac equation generalizes 
Schr\"odinger's equation [3,16]. The potentials $\Delta(r,r')$ and $A_\mu(r)$ 
in Eqs.~(2) and (3) are to be regarded
as effective potentials incorporating the electron-electron interaction
in the spirit of mean-field [11] or density-functional [12] theory.

The pairing of time-conjugate pairs, described by the matrix $\hat{\eta}_0$,
is, in fact, characteristic only of BCS-like spin-singlet superconductors. 
For triplet superconductors, even in
the nonrelativistic case, one needs parity conjugation $\hat{p}$ in addition to
time reversal $\hat{t}$, in order to express the order parameter and the 
underlying pairing state in terms of fundamental discrete symmetries. From 
the two operations, $\hat{p}$ and $\hat{t}$, acting on two-component
(Pauli) spinors one can form four antisymmetric pairing states,
representing singlets and triplets [6,13]. 
The relativistic generalization, $\hat{P}$ and $\hat{T}$, of the symmetry 
operations $\hat{p}$ and $\hat{t}$ is, of course, well known [3,16] and the 
above nonrelativistic prescription for forming pair states is easily
generalized to the relativistic case [6]. Apart from $\hat{P}$ and $\hat{T}$, 
relativity also provides a third fundamental discrete symmetry, charge 
conjugation $\hat{C}$, which can, as a matter of principle, also be used to 
form pair states. From the three operations $\hat{P}$, $\hat{T}$, and 
$\hat{C}$, acting on four-component Dirac spinors, one can form sixteen 
antisymmetric pair states, which are nothing but the sixteen combinations
entering the five superconducting bilinear covariants, obtained independently
above, from considerations of Lorentz covariance.

The Dirac-Bogolubov-de~Gennes equations incorporating all these OP read [17]
$$
\hat{\gamma}^0
[c \vec{\gamma} \cdot \vec{p} + mc^2(1-\hat{\gamma}^0)+q\gamma^\mu A_\mu]
u_{n}(r)
- \sum_{i=0}^{15}
\int d^3r'\Delta_i(r,r')\hat{\eta}_i^\dagger v_{n}(r')
= E_{n}  u_{n}(r)
\eqno(4)
$$
$$
-\hat{\gamma}^0
[c \vec{\gamma} \cdot \vec{p} + mc^2(1-\hat{\gamma}^0)+q\gamma^\mu A_\mu]^*
v_{n}(r)
+ \sum_{i=0}^{15}
\int d^3r' \Delta_i^*(r,r')\hat{\eta}_i^T v_{n}(r')
= E_{n} v_{n}(r),
\eqno(5)
$$
where the sums are over all sixteen matrices $\hat{\eta}_i$, and the
corresponding pair potentials $\Delta_i$. These equations thus contain 
all five possible 
OP compatible with Lorentz covariance. In practice one expects any 
given interaction leading to superconductivity to produce only one of these
covariant OP. The DBdGE for each of these are easily obtained 
from Eqs.~(4) and (5) by setting all pair potentials $\Delta_i$, except for 
those belonging to the IR under consideration, equal to zero.
(In the case of the Lorentz scalar one recovers Eqs.~(2) and (3) in this way.)

\topinsert
\input psfig.sty
\centerline{\hskip10mm\psfig{figure=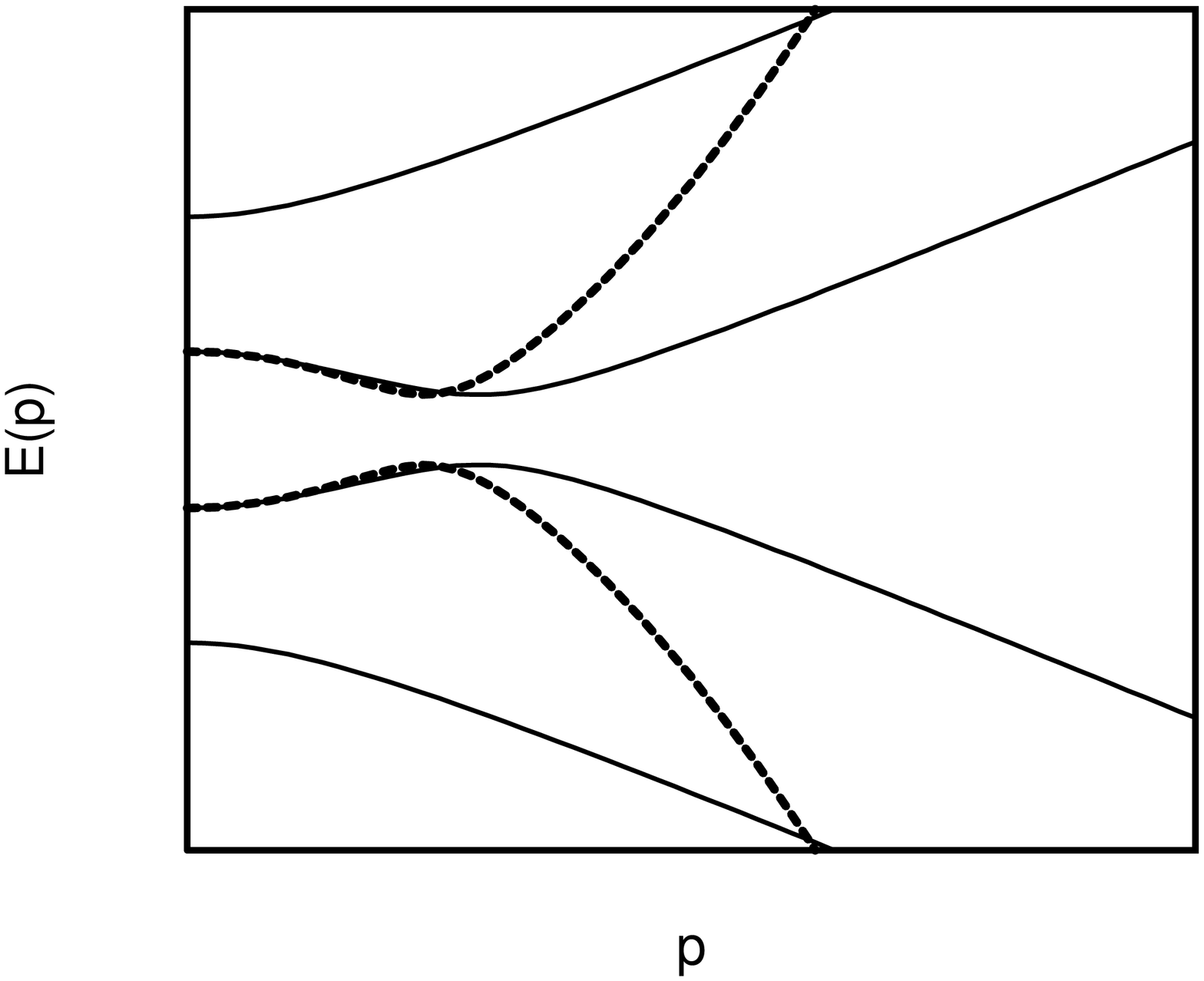,height=9truecm,width=12truecm,angle=270}}
\vskip 0.5truecm
{\bf Figure 1.}
{\it Solid lines:} Schematic energy spectrum for a relativistic BCS-like 
superconductor with a point-contact pair potential, in a constant external 
potential (i.e., disregarding lattice and band structure effects),
calculated by 
analytical diagonalization of the DBdGE~(2) and (3). 
In order to make the superconducting gap and the relativistic gap
visible on the same scale, $\Delta_0$ and $\mu$ were artificially enlarged.
{\it Dashed lines:} Conventional nonrelativistic BCS energy spectrum for the 
same values of $\Delta_0$ and $\mu$. 
More details are given in the main text.
\vskip 12truept
\endinsert

Eqs.~(2) and (3), can be 
diagonalized analytically [5] for spatially uniform
superconductors, employing a point-contact pair potential
$\Delta(r,r')= \Delta_0 \delta(r-r')$ and setting $A_\mu=(v,0,0,0)$, where 
$\Delta_0$ and $v$ are constant in space. The resulting energy spectrum is 
displayed in Fig.~1, together with its nonrelativistic limit, the conventional
BCS spectrum for the same choice of $\Delta(r,r')$ and $\mu$.
For small momenta and energies the relativistic and nonrelativistic spectra
are very similar, but the relativistic superconducting gap is slightly shifted 
to higher momenta with respect to the nonrelativistic one. An analytical 
expression for this shift was obtained in Ref.~[5]. For larger momenta and 
energies the relativistic spectrum aquires features of a free particle solution
of the Dirac equation, such as a linear dispersion relation.

\vskip 28 truept
\centerline{\bf 3.  PAULI EQUATION FOR SUPERCONDUCTORS}
\vskip 12 truept

Although the task to set up a relativistic generalization of the 
Bogolubov-de~Gennes formulation of the BCS theory is formally completed with 
the derivation of the Dirac-Bogolubov-de~Gennes equations and all possible
order parameters they may contain, the underlying physics comes out much 
clearer by leaving the level of the Dirac equation and proceeding to the 
weakly relativistic and the nonrelativistic limits. 
These limits are important because
(i) the nonrelativistic limit of the DBdGE must be the conventional BdGE, hence
recovering this limit constitutes an essential test of the theory, and
(ii) in the weakly relativistic limit (defined as including terms up to second
order in $v/c$, where $v$ is a typical particle velocity) one obtains the
first relativistic corrections to the conventional theory of superconductivity.

By systematic elimination of the lower components of the DBdGE (i.e., those 
which are suppressed by factors of $v/c$ in the weakly relativistic limit) one
obtains from Eqs.~(2) and (3) the pair of equations [5,7]
$$
\left[h(r)+\delta h(r)\right] u_n(r) 
+ \int d^3r' \left[\Delta(r,r')i\hat{\sigma}_y + \delta\Delta(r,r')\right]
v_n(r') 
= E_n u_n(r)
\eqno(6)
$$
$$
-\left[h(r)+\delta h(r)\right]^* v_n(r) -
\int d^3r'\left[\Delta^*(r',r) i\hat{\sigma}_y-\delta\Delta^\dagger(r',r)
\right]
u_n(r')
= E_n v_n(r)
\eqno(7)
$$
where $u_n$ and $v_n$ are two-component (Pauli) spinors, 
$h(r)=\frac{1}{2m}[\vec{p}-\frac{q}{c}\vec{A}(r)]^2
+v(r) - \mu \vec{\sigma}\vec{B}(r)$ is the normal-state Hamiltonian,
including the vector potential $\vec{A}$ and the magnetic field $\vec{B}$,
and $\mu=\hbar q/(2mc)$ is the Bohr magneton.
Neglecting the relativistic correction terms $\delta h$ and $\delta \Delta$, 
one obtains from Eqs.~(6) and (7) the traditional BdGE
for spin-singlet superconductors [11]. The term
$$
\delta h(r) = \frac{1}{4m^2c^2}\left[
\hbar \vec{\sigma}\cdot[\nabla v(r)]\times \vec{p} 
+ \frac{\hbar^2}{2}\nabla^2 v(r)
- \frac{p^4}{2m} \right]
\eqno(8)
$$
represents the second-order relativistic corrections appearing already in the
normal state, i.e., spin-orbit coupling, which involves the gradient of the 
lattice potential, the Darwin term, containing second derivatives of the
lattice potential, and the mass-velocity correction, which is quartic in $p$.
Similarly,
$$
\delta \Delta(r,r')=  \frac{1}{4m^2c^2}\left[
\hbar \vec{\sigma}\cdot\left[\nabla\Delta(r,r')\right]
\times \vec{p}'
+\frac{\hbar^2}{2} \left(\nabla+\nabla'\right)^2
\Delta(r,r') 
\right] i\hat{\sigma}_y
\eqno(9)
$$ 
contains the second-order relativistic corrections involving the pair 
potential, i.e., those which appear only in the superconducting state.
(The prime on $p$ and $\nabla$ denotes a derivative with repect to the primed
coordinate.)
These terms depend on the pair potential in a similar way in which
those of $\delta h$ depend on the lattice potential, and are referred to as
the {\it anomalous} spin-orbit coupling (ASOC) and anomalous Darwin terms, 
respectively.

More correction terms arise if additionally triplet OP are included. 
In this case already in the nonrelativistic (zero order in $1/c$) terms in 
Eqs.~(5) and (6) one must add to the term $\Delta(r,r')i\hat{\sigma}_y$ 
the sum $\sum_{j=x,y,z} \Delta_j(r,r') \hat{\sigma}_j\hat{\sigma}_y$ [14,15].  
Furthermore, Eqs.~(8) and (9) stem from Eqs.~(2) and (3), and hence do
not include OP involving charge conjugation, neither for singlets nor for
triplets.
The generalization of Eqs.~(8) and (9) involving {\it all} possible OP 
discussed in the previous sections is found from Eqs.~(4) and (5) [17]
and reads
$$
\delta h(r) = \frac{1}{4m^2c^2}\left[
\frac{\hbar^2}{2}\nabla^2 v(r) +
\hbar \vec{\sigma}\cdot(\nabla v)\times \vec{p} -\frac{p^4}{2m}
+\frac{1}{2mc^2}\hat{\cal D}_{12}\hat{\cal D}_{21}^* 
\right]
\eqno(10)
$$
and
$$
\delta \Delta (r,r') = 
\frac{1}{2mc}\left[
\vec{\sigma}\left(\vec{p}-\frac{q}{c}\vec{A}(r)\right) \hat{\cal D}_{21} 
+\hat{\cal D}_{12}
\vec{\sigma}^*\left(\vec{p}'-\frac{q}{c}\vec{A}(r')\right)^*\right]
$$
$$
+\frac{1}{4m^2c^2}\left[
\vec{\sigma}\vec{p}\hat{\cal D}_{22} \vec{\sigma}^*\vec{p}'^*
+\frac{\hbar^2}{2}\left(\nabla^2 \hat{\cal D}_{11} 
+ \hat{\cal D}_{11} \nabla'^2 \right)
\right],
\eqno(11)
$$
where the $4\times4$ matrix $\hat{\cal D}$ is defined in terms of the complete
set of $\hat{\eta}$ matrices as
$$
\hat{\cal D} = -\sum_{i=0}^{15} \Delta_i(r',r) \eta_i^\dagger,
\eqno(12)
$$
and the action of the product $\hat{\cal D}_{12}\hat{\cal D}_{21}^*$ 
on a function $f(r)$ should be interpreted as
$$
\hat{\cal D}_{12}\hat{\cal D}_{21}^* f(r) =
  \sum_{i,j=0}^{15} \int dr''\Delta_j(r'',r) \Delta^*_i(r',r'') f(r')
  \eta_{j 12}^\dagger \eta_{i 21}^T.
\eqno(13)
$$
The subscripts in, e.g., $\hat{\cal D}_{12}$, refer to $2\times2$ 
submatrices, i.e., $\hat{\cal D}_{12}$ is the upper right block of 
$\hat{\cal D}$. 
If all pair potentials but that multiplying $\hat{\eta}_0$ are set equal to 
zero (i.e., if one specializes to a BCS-like spin-singlet OP) these equations 
reduce to (8) and (9). On the other hand, through the OP matrices 
$\hat{\eta}_1 \ldots \hat{\eta}_{15}$ and the corresponding pair potentials
they contain all relativistic corrections for singlet and triplet
superconductors up to second order in $1/c$, including those forming their 
Cooper pairs via charge conjugation.

We now turn to a discussion of the physics behind the weakly relativistic
correction terms of Eqs. (8), (9), (10), and (11).
We start by considering only the corrections appearing for BCS-like singlet 
superconductors, i.e., those given in Eqs.~(8) and (9).
The spin-orbit, mass-velocity, and Darwin corrections to $h(r)$, contained
in Eq.~(8), appear already in the normal state and are well known from
other areas of physics [3,4,16]. Although these corrections are thus not of
superconducting origin, their effect on observables in a superconductor can be
dramatically modified by superconducting coherence. The conventional spin-orbit 
coupling (SOC) term in Eq.~(8), for example, breaks rotational symmetry
of the spin degrees of freedom with respect to the orbital degrees of freedom.
It is well known that this broken symmetry gives rise to circular dichroism in
the magneto-optical response of metals [4]. Below the critical temperature
$T_c$ the dichroic response of a superconductor is drastically modified. 
A microscopic theory of this modification was 
recently worked out in Refs.~[18] and [19] on the basis of Eqs.~(6) to (9),
and the results can qualitatively account for experimental observations.
As a representative result we present in Fig.~2 a graph displaying 
the circular dichroic response of a simple model superconductor as a function
of temperature, normalized by dividing by the corresponding normal-state
result [19]. Without SOC both numerator and denominator would be zero, 
while without superconducting coherence the curve would be flat throughout.
The strong peak seen immediately below the transition temperature thus 
arises only from the simultaneous presence of superconducting coherence
and SOC [19]. 

\topinsert
\input psfig.sty
\centerline{\hskip10mm\psfig{figure=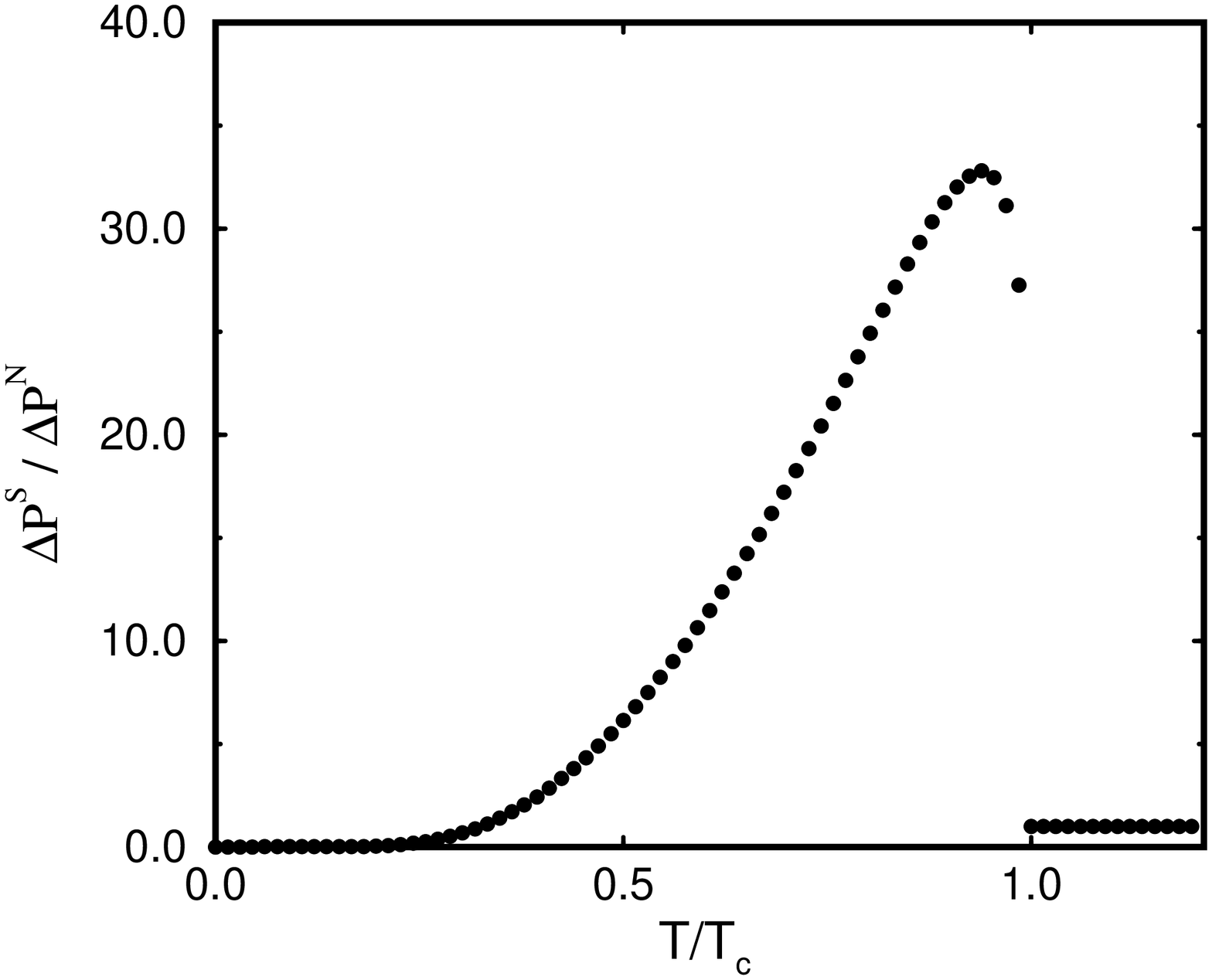,height=9truecm,width=12truecm,angle=270}}
{\bf Figure 2.}
Difference in power absorption $\Delta P^S$ between left-hand circularly
polarized light and right-hand circularly polarized light in the
superconducting state, divided by the corresponding normal-state difference
$\Delta P^N$, as a function of temperature divided by the transition
temperature $T_c$. The difference $\Delta P^S$ is proportional to
$\Im [\hat{\sigma}^S_{xy}(\omega,T,H)]$, the imaginary part of the offdiagonal
elements of the conductivity tensor in the superconductor, as a function of
frequency, temperature and magnetic field ($\omega = 4.5 meV$ and $H=0.05 T$
for the data in the figure) [19].
\endinsert

The mass-velocity term, on the other hand, 
provides a small correction to the bare electron mass. 
It attests to the ingenuity of experimental physicists that the very small
change of the Cooper pair mass in a superconductor due to the relativistic
mass enhancement has already been measured [20,21,22]. 
Another consequence of this term in superconductors is the small shift of
the energy gap of a homogeneous superconductor, mentioned at the end 
of Section 2.

The anomalous spin-orbit and Darwin corrections to $\Delta(r,r')$, given
in Eq.~(9), are fundamentally different from those to $\delta h(r)$, since 
they depend explicitly on the pair potential and are nonzero only in the 
superconducting state of matter. These terms were derived for the first time
in Ref.~[5]. The anomalous spin-orbit coupling provides a contribution to 
dichroism in superconductors, which can be distinguished from that of 
conventional SOC due to their very different temperature dependence [18,19].        
The first appearence of an SOC term containing the pair potential, i.e., the
term nowadays called ASOC, dates back to 1985, when Ueda and Rice
postulated such a term on group theoretical grounds in their
phenomenological treatment of p-wave superconductivity [23]. However, at 
that time it was not clear how such a term could be obtained
microscopically, and what its detailed form was. 
These questions were answered only ten years
later on the basis of the theory outlined above [5]. 
Both Darwin terms, the conventional and the anomalous one, have also
been rederived phenomenologically [9].
This rederivation showed that the anomalous Darwin term 
can be understood as a consequence of relativistic fluctuations of paired
particles in the pair potential of the superconductor, in a similar
way in which the conventional Darwin term can be understood as a
consequence of fluctuations of charged particles in the electric (lattice)
potential [16].

We now turn to the relativistic corrections appearing for more exotic
superconductors, contained in Eqs. (10) and (11).
A closer look at these equations reveals that there are several
terms which are straightforward generalizations of those in Eqs.~(8) and (9),
but also two terms which do not have counterparts in those equations.
The first of these is the term containing the pair potential operator
in $\delta h(r)$. It is a new feature of these equations, not present in
other Bogolubov-de~Gennes equations, that terms related to
superconductivity appear also on the diagonal of the BdG matrix.
The second is the cross term in $\delta \Delta(r,r')$, containing both the 
pair potential operator and the vector potential. 
Although this latter term has the usual minimal coupling form (in which 
$\vec{p}$ is replaced by $\vec{p}-(q/c)\vec{A}$) it is, in fact, highly 
unusual, since it leads to a term containing 
the pair potential multiplied by the vector potential. 
It is of considerable fundamental interest that such unusual coupling
becomes possible in the theory of superconductivity, but detailed consequences
of this remain to be explored.
Both of these novel terms contain the offdiagonal blocks of the matrix
$\hat{\cal D}$, which are nonzero only for OP formed with relativistic
charge conjugation as fundamental discrete symmetry. These OP do vanish
in the strictly nonrelativistic limit [6,7,17], and are thus
intrinsically relativistic. 
Interestingly, for these OP there are no anomalous Darwin and spin-orbit
terms, since these involve only the diagonal blocks of the matrix
$\hat{\cal D}$, which are zero for the OP involving charge conjugation.
In the absence of this purely relativistic type of pairing the
matrix $\hat{\cal D}$ is block diagonal, and all terms containing
$\hat{\cal D}_{12}$ and $\hat{\cal D}_{21}$ vanish. The remaining terms
provide the leading order relativistic corrections to the nonrelativistic
singlet [11] and triplet [14] Bogolubov-de~Gennes equations.

To conclude this section we emphasize that all results up to this point are
independent of the form of the interaction leading to superconductivity.
The role of the interaction is to select among the {\it possible}
OP, discussed in Section 1, the one which is {\it realized} in a given
superconductor. Relativity alone does not determine what interactions
can give rise to superconductivity, but it puts strong constraints on the
form and transformation behaviour of the resulting OP (e.g., it limits their
number to sixteen, forming the five covariant combinations listed above). 
Furthermore, regardless of the detailed nature of the interaction, if there 
is superconductivity at all, then there will be relativistic corrections to it,
and the form of these is universal, i.e, independent of the interaction 
itself. For example, any superconductor with a BCS-like singlet OP, no matter
what (quasi)particles  are paired and by what interaction, must display the
relativistic corrections contained in Eqs.~(8) and (9).

\vskip 28 truept
\centerline{\bf 4.  EXPERIMENTAL ASPECTS AND OUTLOOK}
\vskip 12 truept

By now there are several predictions for experiment, which have been extracted 
from the general theory outlined above. These are
(i) the existence of new types of order parameters in a relativistic theory,
which vanish in the nonrelativistic limit [6,7,17], 
(ii) a small shift in the position of the energy gap [5],
(iii) relativistic fluctuations in the pair potential, giving rise to the 
anomalous Darwin term [9],
(iv) a drastic modification of the contribution of SOC to the dichroic 
response of superconductors below $T_c$ [18,19],
(v) a contribution of ASOC to the same dichroic response [18,19],
(vi) the influence of relativity on the OP symmetry, even for conventional
superconductors [5,6,7], and
(vii) the relativistic change of the Cooper pair mass [5,7,20,21,22].
Clearly, the task to develop the theory to the point at which meaningful
comparison with experiment can be made is not an easy one, and only first
(preliminary, but encouraging) steps could be reported in this chapter.
However, a few general considerations may facilitate experimental detection 
of these effects: 

First of all, since relativistic effects become important at high velocities,
one should study superconductors in which the electrons move rapidly.
This is the case either if the normal-state metal has a very high
Fermi velocity, or if there are sufficiently heavy elements in the lattice,
whose orbitals are hybridized with the conduction band.
Examples of such systems are the heavy-fermion superconductors like
$UPt_3$ -- containing some of the heaviest elements found in solid-state
compounds, the high-temperature superconductors -- which include between the
$CuO$ layers heavy elements like $Ba$, $La$ or $Hg$, or BCS superconductors
with heavy atoms in the lattice, such as $Pb$. 
Apart from high velocities, rapid spatial variations of the pair potential
particularly favour the anomalous relativistic effects (ASOC and the anomalous 
Darwin term), since these depend on derivatives of the pair potential.
Situations in which the pair potential varies rapidly are, e.g., 
superconductivity in thin films and small grains, multilayers, and the vortex
state. In all these cases a short coherence length allows more rapid spatial
variations, and thus increases the anomalous relativistic effects (an
observation which points again at high-temperature superconductors).
Presumably, the most important relativistic corrections in these materials are 
those of Eq.~(8) (which are already routinely taken into account
in relativistic band-structure codes for the normal state, but not included
in most approaches to superconductivity). However, new and interesting
phenomena can be expected to arise as a consequence of the anomalous
corrections of Eq.~(9).

A more detailed exploration of the fascinating territory of the interplay of
superconducting coherence and relativistic covariance remains a task for the
future, but already now it seems certain that the two areas can benefit
from each other both in the study of down-to-earth superconductors, which was
the subject of this paper, and in the field of cosmological and
high-energy analogies of coherence and superconductivity [24], to which
many of our results also apply.

\vskip 21 truept
\centerline{\bf ACKNOWLEDGMENTS}
\vskip 9 truept
KC thanks the FAPESP for support under grant JP 1999/09269-3.
All three of us thank B.~L. Gy\"orffy for many useful discussions.
This work was supported in part by the Deutsche Forschungsgemeinschaft.

\vskip 21 truept
\centerline{\bf REFERENCES}
\vskip 9 truept

\item{[1]} {\it The Effects of Relativity in Atoms, Molecules and the Solid 
State}, eds. S.~Wilson, I.~P. Grant, B.~L. Gyorffy (Plenum, New York, 1991).
\item{[2]} {\it Relativistic Effects in Atoms, Molecules and Solids},
NATO ASI Series B87, ed. G.~L. Malli (Plenum, New York, 1983).
\item{[3]} P.~Strange, {\it Relativistic Quantum Mechanics with Applications 
in Condensed Matter and Atomic Physics}
(Cambridge University Press, Cambridge, 1998).
\item{[4]} H.~Ebert, {\it Rep. Prog. Phys.} {\bf 59} 1665 (1996).
\item{[5]} K.~Capelle and E.~K.~U.~Gross, {\it Phys. Lett. A} {\bf 198} 261
(1995),
\item{[6]} K.~Capelle and E.~K.~U.~Gross, {\it Phys. Rev. B} {\bf 59} 7140
(1999).
\item{[7]} K.~Capelle and E.~K.~U.~Gross, {\it Phys. Rev. B} {\bf 59} 7155
(1999).
\item{[8]} M.~Marques, K.~Capelle, and E.~K.~U.~Gross, {\it Physica C} 
{\bf 317-318} 508 (1999).
\item{[9]} K.~Capelle, {\it Phys. Rev. B}, in press (to appear in 2001).
\item{[10]} J.~Bardeen, L.~N.~Cooper and J.~R.~Schrieffer, Phys. Rev. {\bf 108}
1175 (1957).
\item{[11]} P.~G.~de~Gennes, {\it Superconductivity of Metals and Alloys}
(Adison Wesley, Reading, 1966).
\item{[12]} L.~N. Oliveira, E.~K.~U. Gross, W. Kohn,
Phys. Rev. Lett. {\bf 60} 2430 (1988).
\item{[13]} P.~W. Anderson, Phys. Rev. B {\bf 30} 4000 (1984).
\item{[14]} K.~Capelle and E.~K.~U.~Gross, {\it Int. J. Quantum Chem.}
{\bf 61} 325 (1997).
\item{[15]} D.~Vollhardt and P.~W\"olfle, {\it The Superfluid Phases of Helium
3} (Taylor \& Francis, London, 1990).
\item{[16]} J.~J.~Sakurai, {\it Advanced Quantum Mechanics}
(Adison Wesley, Reading, 1967).
\item{[17]} M.~Marques, Ph.D. thesis, University of W\"urzburg (2000).
\item{[18]} K.~Capelle, E.~K.~U.~Gross, and B.~L.~Gyorffy,
{\it Phys. Rev. Lett.} {\bf 78} 1872 (1997).
\item{[19]} K.~Capelle, E.~K~.U.~Gross, and B.~L.~Gyorffy,
{\it Phys. Rev. B} {\bf 58} 473 (1998).
\item{[20]} J.~Tate, B.~Cabrera, and S.~B.~Felch, {\it Phys. Rev. Lett.}
{\bf 62} 845 (1989).
\item{[21]} B.~Cabrera and M.~E.~Peskin, {\it Phys. Rev. B} {\bf 39} 6425
(1989).
\item{[22]} J.~Tate, S.~B.~Felch and B.~Cabrera, {\it Phys. Rev. B} 
{\bf 42} 7885 (1990).
\item{[23]} K.~Ueda and T.~M.~Rice, {\it Phys. Rev. B} {\bf 31} 7149 (1985).
\item{[24]} G.~E.~Volovik, {\it Superfluid analogies of cosmological
phenomena}, gr-qc/0005091 (2000).

\end